\documentclass[12pt]{article}
\usepackage{amsmath,amsfonts,amssymb,alltt,authblk}
\usepackage{graphicx,color}
\usepackage{epstopdf}
\begin{document}

\baselineskip=7.0mm
\setlength\parindent{0pt}
\newcommand{\be} {\begin{equation}}
\newcommand{\ee} {\end{equation}}
\def\a{\alpha}
\def\b{\beta}
\def\g{\gamma}
\def\d{\delta}
\def\D{\Delta}
\def\t{\tau}
\def\lg{\langle}
\def\rg{\rangle}
\author[]{Marco Oestereich}
\author[]{J\"urgen Gauss}
\author[]{Gregor Diezemann}
\affil[]{Department Chemie, Johannes Gutenberg-Universit\"at Mainz, Duesbergweg 10-14, 55128 Mainz, Germany}
\title
{Markov State Model for the forced unfolding of a small peptide}

\maketitle
\begin{abstract}
\noindent 
In typical single-molecule force spectroscopy experiments the mechanical unfolding of molecular complexes or biomolecules is studied applying a force ramp to one end of the system while the other end is kept fixed in space. 
The computational counterpart of this type of experiments can routinely be performed using molecular dynamics simulations with atomistic resolution.
However, due to the large difference in time scales often coarse graining procedures are applied in the simulations.
Most of the applied techniques do not allow to follow the atomistic details of the relevant conformational transitions due to the structural simplifications used to speed up the simulations.
Here, we apply an earlier developed dynamic coarse graining technique based on Markov state modeling to a model peptidic system that does not unfold in a simple two-state manner.
Using the donor-acceptor distances of the helical hydrogen bonds as collective variables and performing a dimension reduction technique allows us to construct a Markov model of the unfolding process that correctly represents the microscopic behavior of the system.
The chosen example shows that the method can be used to mimick the mechanical unfolding process of systems for which the end-to-end distance does not provide a sufficient order parameter and that do not unfold in a simple cooperative manner.
\end{abstract}
\newpage
\section*{I. Introduction}
Conformational transitions in supermolecular systems like proteins or DNA often take place on vastly different time scale and therefore can only be studied successfully using the combination of a number of experimental techniques\cite{Bieri:1999}.
To speed up slow processes the application of mechanical forces has developed to a standard technique that allows to extract details 
of the energy landscape of the system\cite{Woodside:2014,Bustamante:2020,Li:2022}.
Furthermore, relevant informations about the kinetics of the relevant molecular transformations can be extracted applying specific models of barrier crossing\cite{Hummer:2003,Dudko:2006,Dudko:2008}.

Today, computer simulations are routinely used to investigate atomistic details of structural and dynamical aspects of soft-matter physics\cite{Lee:2009,Luitz:2015,Perilla:2015}.
In analogy to experimental force spectroscopy force probe molecular dynamics (FPMD) simulations, also called steered molecular dynamics simulations, are applied to resolve not only questions regarding the nature and the time scale of conformational transitions but also to perform non-equilibrium free-energy calculations\cite{Isralewitz:2001,Park:2004,Chen:2011}. 
Apart from recent examples\cite{Rico:2019,Valotteau:2019}, a direct comparison with results from experimental force spectroscopy is mainly hampered by the fact that the pulling velocities usually used in the simulations are many orders of magnitude larger than the experimentally accessible values\cite{Franz:2020}.
To overcome these limitations, a number of coarse graining techniques have successfully been applied to this kind of 
simulations\cite{Best:2008,Habibi:2016,Mucksch:2016}.
While most of commonly employed coarse graining methodologies use structural simplifications, approximative forces or implicit solvent techniques, Markov state models (MSMs) can be used to extrapolate short trajectories to longer times under the assumption that the dynamics of the relevant variables is Markovian\cite{Ghosh:2017,G89}.
We have applied the transition state reweighting analysis method (TRAM)\cite{Wu:2016} to treat the time-dependence of the protocol using a constant pulling velocity and were able to fill the gap between the times scales typical for FPMD simulations and experiments\cite{G89}.
We investigated a simple model system consisting of a calixarene catenane dimer undergoing a reversible two-state kinetics between a closed and an open structure and the end-to-end distance served as an order parameter.
However, for more complex systems the assumption of the existence of a single order parameter describing the mechanical (un)folding process is not necessarily valid. 
For small peptides, it has been shown that the end-to-end distance is an insufficient order parameter under the application of an external force\cite{Knoch:2017}.
Therefore, in the present work, we investigate the applicability of an extension of the method used for the simple calixarene system to a $\b$-peptide exhibiting a more complex mechanical unfolding pathwas via an intermediate state\cite{G84}.

The remainder of the paper is organized as follows. 
In the next section, we discuss the computational methodologies and in Section III the results are presented and discussed.
The paper closes with some conclusions in section IV.
\section*{II. Methods and Results}
\begin{figure}[h!]
\vspace{-0.25cm}
\centering
\includegraphics[width=12.0cm]{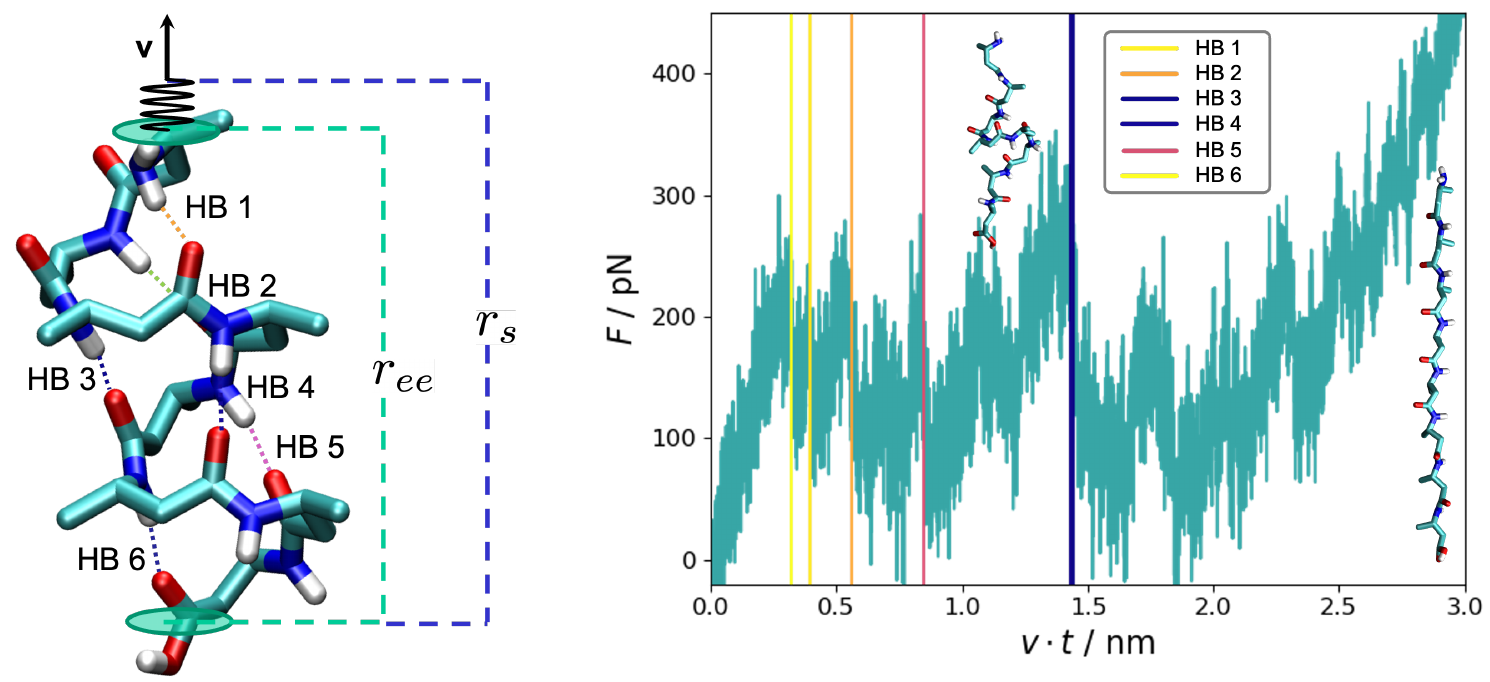}
\vspace{-0.5cm}
\caption{Left: Structure of the $3_{14}$-helix formed by the $\b$-alanine octamer with the H-bonds indicated and enumerated starting at the N-terminus. 
The dashed lines indicates the definition of the end-to-end distance $r_{ee}$ (green) and the sampling distance $r_s$ (blue).
The arrow indicates the pulling direction determined by the vector ${\vec v}$.
\newline
Right: Force versus extension curve obtained from a FPMD simulation at $T=240$ K for a loading rate $\mu=K\cdot v=1$ N/m. The vertical lines indicate the times of the breaking of the respective H-bond as indicated by the color code.
The structures shown correspond to the intermediate structure and the fully stretched conformation.
}
\label{Fig1}
\end{figure}

Our model system consists of a $\b$-alanine octamer solved in methanol.
This system has been studied in various situations, in equilibrium\cite{Cheng:2001}, under the impact of a constant force\cite{Knoch:2017} and also using a time-varying force in atomistic FPMD simulations\cite{G84,G85} and also in FPMD simulations employing an adaptive resolution method to allow for coarse graining the solvent degrees of freedom\cite{Praprotnik:2006,G104}. 
In constrast to $\a$-peptides the structurally slightly different $\b$-peptides form stable helical structures  also for small numbers of residues\cite{Seebach:1997,Gademann:1999}.
The equilibrium $3_{14}$-helix formed by the $\b$-alanine octamer is stabilized by 6 hydrogen bonds (H-bonds) and the end-to-end distance is $r_{ee}\simeq1.2$ nm while one has $r_{ee}\simeq3.5$ nm in the stretched elongated conformation, cf. Fig.\ref{Fig1}.
We note that in case of FPMD simulations $r_{ee}$ is always to be identified with the projection of the vector connecting the carbon atom of the C-terminus with the nitrogen atom of the N-terminus, $\vec{r}_N-\vec{r}_C$, onto the pulling direction.
The mechanical unfolding process of the $\b$-alanine octamer proceeds via an intermediate structure with the innermost H-bond still intact while the outer ones are broken first, cf. Fig.\ref{Fig1}.
Thus, the unfolding of the $\b$-alanine octamer is more complicated than that of a simple two-state folder because 
one has a two-step unfolding scenario (three states) and furthermore the end-to-end distance does not represent a good order parameter, rendering the theoretical treatment more challenging.
\subsubsection*{FPMD simulations}
All atomistic simulations were performed using the GROMACS 2020.1 program package\cite{Abraham:2015} using a stochastic dynamics integrator\cite{Goga:2012} at T=240 K with a friction constant of 0.2 ps. 
For the $\b$-alanine octamer, the atomistic GROMOS 53a6 force field\cite{Oostenbrink:2004} and for ethanol the MeOH model B3 force field\cite{Walser:2000} were used. 
For the short-range interactions, a cut-off of 1.2 nm was used and the particle mesh Ewald summation method was applied to compute the long-range Coulomb interactions\cite{ewaldsum,ewaldsum2}.
All bonds containg hydrogen were constrained to their equilibrium bond length using the LINCS algorithm\cite{Hess:1997}
allowing for a time step of 2 fs. 
The long-range corrections for the energy and the pressure were computed using a dispersion correction for the long-range van der Waals interactions\cite{Allen:1987,Abraham:2018}.
The neighbor list was updated after 25 simulation steps. 
We used Cartesian periodic boundary conditions in all simulations.
The cubic simulation box of length $5.50\times3.45\times3.45$ nm was filled with one $\b$-alanine octamer and 1027 methanol molecules.
After an equilibration period of 1 ns, the system was equilibrated at 240 K and the pressure was set to 1 bar using a Berendsen barostat\cite{Berendsen:1984} with a time constant of 0.5 ps and a compressibility of $1.248\cdot10^{-4}$ bar$^{-1}$ for another 1 ns.
The resulting box dimensions were hardly changed and all production run simulations were performed in the canonical ensemble.

For the FPMD simulations, the carbon atom at the C-terminus acted as the reference group of the peptide that was fixed in space and a harmonic potential was applied to the nitrogen atom of the N-terminus (pulled group) and for the pulling geometry the direction periodic algorithm was employed.
Since these simulations were performed in the so-called force ramp mode with a constant pulling velocity, the force measured at the position of the spring reads:
\be\label{Fpull.def}
F(t)=K(v\cdot t-x_P(t)) = K(r_s(t)-r_{ee}(t)).
\ee
Here, $K$ denotes the spring constant of the harmonic potential applied, chosen as $K=1$ N/m, $v$ is the pulling velocity,
$x=v\cdot t$ is the extension, and $x_P(t)$ is the deviation of the position of the pulled group from its initial value, $x_P(0)$.
In the second equality, the definitions of the end-to-end distance $r_{ee}$ and the sampling distance $r_s$ as the distance between the spring position and the reference group, are used, cf. Fig.\ref{Fig1}.
The system was always pulled until the maximum extension of 4.2 nm was reached.
A typical example for a force versus extension curve (FEC) is shown in Fig.\ref{Fig1}.
\subsubsection*{Markov State Modeling}
In order to construct a MSM for the forced unfolding process of the $\b$-alanine octamer, in a first step we had to choose a meaningful set of collective variables (CVs).
As mentioned above, the end-to-end distance $r_{ee}$ provides only an insufficient reaction coordinate in the presence of an external force.
Furthermore, it turned out that the unfolding kinetics of the peptide at $T=298$ K is not Markovian when $r_{ee}$ is taken as the relevant CV.
One order parameter that often is used in the treatment of biomolecules is the fraction of native constants and we have also used this quantity successfully in the context of FPMD simulations\cite{G104}.
In general, the choice of the relevant CVs depends on the system and the context and there does not appear to exist a general recipe for this choice\cite{Noe:2017,Bhakat:2022}.
Since it is well known that the breaking of the H-bond network is intimately related to the unfolding process of peptides and the average fraction of native contacts is closely related to the number of H-bonds, in the present work we use the donor-acceptor distance of the 6 H-bonds as CVs, cf. Fig.\ref{Fig1}.
In order to restrict ourselves to the slowest relevant time scales in the dynamics of the folding kinetics, a dimension reduction using the time-lagged independent component analysis (TICA) was applied\cite{Molgedey:1994,PerezHernandez:2013}. 
We chose a lag time of $\t_{TICA}$=1.5 ns for the equilibirum simulations at T=298 K and $\t_{TICA}$=6.0 ns for the FPMD simulations at T=240 K because for this choice the slowest implied time TICA scale is independent of $\t_{TICA}$. 
We found that using the three most relevant independent components (ICs) yielded more than 95 \% of the cumulative kinetic 
variance\cite{Noe:2015} and therefore used these in addition to $r_{ee}$ as relevant order parameters for the kinetics.
In accord with the findings by Knoch and Speck\cite{Knoch:2017} we observe that the dynamics is Markovian in equilibrium at $T=298$ K and also under the influence of an external force at $T=240$ K if one considers IC$_1$, IC$_2$, IC$_3$ and $r_{ee}$.
Using k-means clustering\cite{MSMbook}, the reduced four-dimensional configuration space is discetized into 500 states (cluster centers)using the PyEMMA software package\cite{Scherer:2015}.

\begin{figure}[h!]
\vspace{-0.25cm}
\centering
\includegraphics[width=10.0cm]{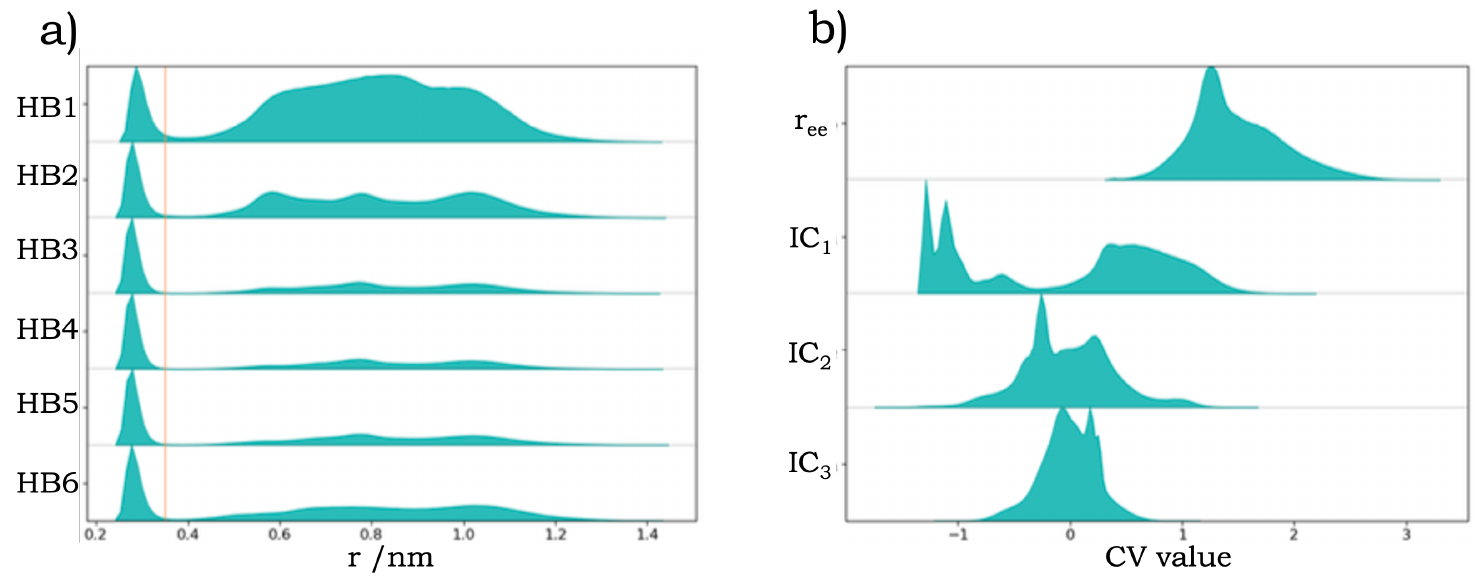}
\includegraphics[width=8.0cm]{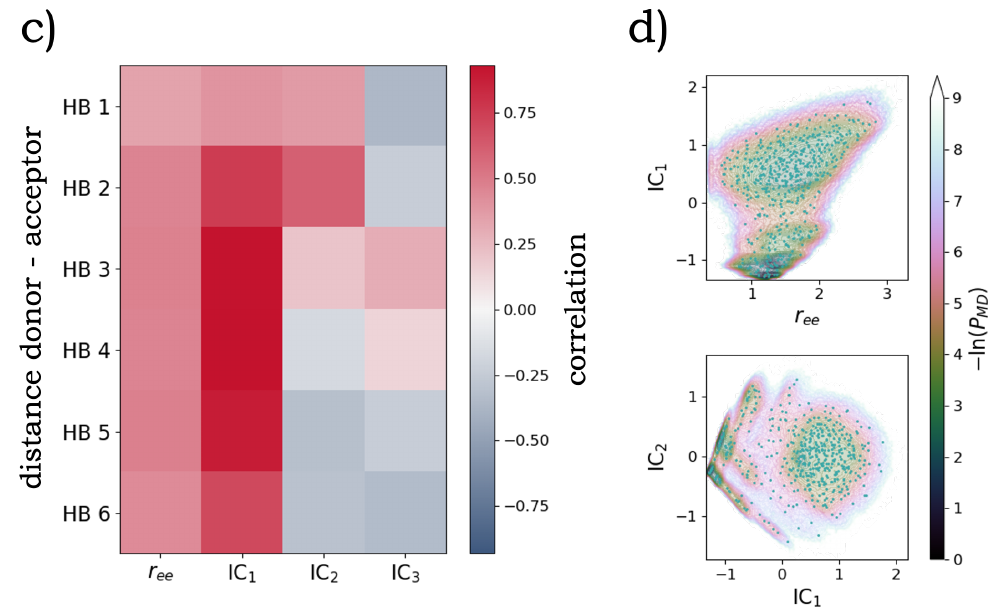}
\vspace{-0.5cm}
\caption{a): Distributions of donor-acceptor distances for the H-bonds obtained from a 1 $\mu$s simulation at T=298 K. For distances larger than 0.35 nm (horizontal line) the H-bond can be considered as open.
\newline
b): Distribution of the values of the CVs for the same simulation.
\newline
c): Correlation diagram for the H-bonds with IC$_1$, IC$_2$, $r_{ee}$ at $T=298$ K.
\newline
d): Potential of mean force ($-\ln{(P_{MD})}$) with $P_{MD}$ denoting the probability of finding a given pair of values in an equilibrium MD simulation at $T=298$ K.
The circles represent the cluster centers of the discretization of phase space.
}
\label{Fig2}
\end{figure}

In Fig.\ref{Fig2}a) we show the distribution of donor-acceptor distances as a function of the distance itself resulting from 25  equilibrium MD simulations, each of 1 $\mu$s duration, at 298 K starting from the 14-helix.
Additionally, the distributions of the CVs (Fig.\ref{Fig2}b)) and the correlation diagram of the donor-acceptor distances
with the CVs (Fig.\ref{Fig2}c)) are presented for the same simulations.
From the correlation diagram it can be observed that IC$_1$ behaves similar to $r_{ee}$ and both quantities take on large values when all H-bonds are opened (donor-acceptor distance larger than 0.35 nm).
This means that the helical structure corresponds to a small value of $r_{ee}$ and a negative value of IC$_1$.
Furthermore, IC$_2$ shows different correlations with the H-bonds located at the N-terminus (positive) and at the C-terminus (negative) showing that a positive value of IC$_2$ indicates open H-bonds HB1 and HB2 which means the peptide is opened at the N-terminus.
Likewise a negative value of IC$_2$ indicates an open C-terminus.
Furthermore, negative values for IC$_3$ indicate the conformations of the intermediate that has been observed in the earlier FPMD simulations, with open endstanding loops and helical structure in the center of the peptide.
The distributions of the CVs show a substantial fraction of helical conformations.

The potential of mean force (PMF) presented in Fig.\ref{Fig2}d) contains similar information.
In the upper plot the minimum in the PMF located at small $r_{ee}$ and negative IC$_1$ clearly shows the helical structure and the broad minimum for positive IC$_1$ corresponds to a number of partly unfolded structures.
From this plot it also becomes evident why $r_{ee}$ alone is an insufficient order parameter because quite different partially unfolded structures (varying IC$_1$) correspond to the same value of $r_{ee}$.
In the lower part, the folded structure is found at the minimum located at negative IC$_1$ and nearly vanishing IC$_2$ and also here partially folded structures are located in the area of the minimum found at positive IC$_1$ and varying IC$_2$.

\begin{figure}[h!]
\vspace{-0.25cm}
\centering
\includegraphics[width=8.0cm]{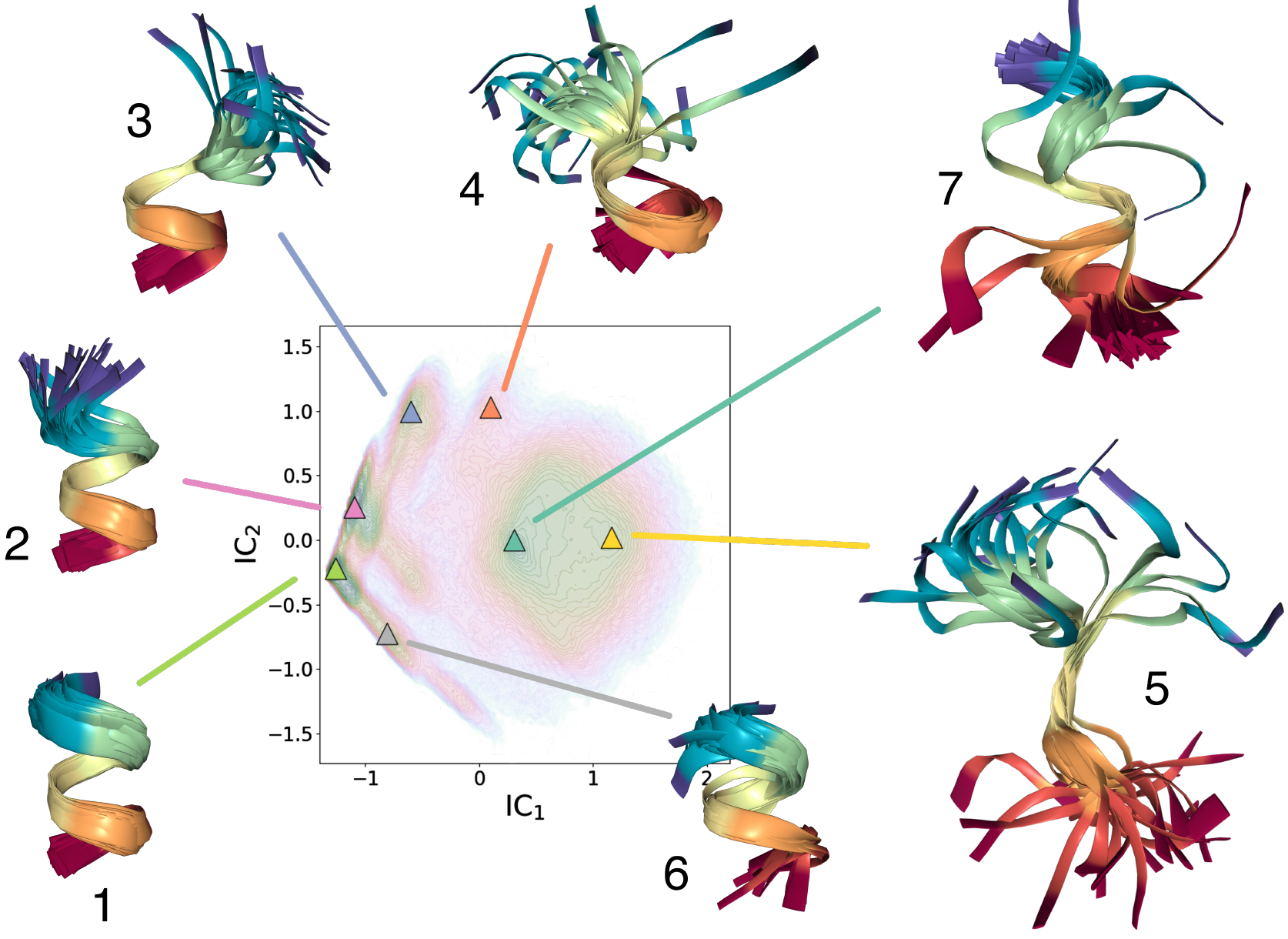}
\vspace{-0.25cm}
\caption{Thermal unfolding pathway of the $\b$-alanine octamer for T=298 K. The unfolding proceeds from the N-terminus (blue) to the C-terminus (red). Also shown is a misfolded structure (7) and one opened partially at the C-terminus (6). 
}
\label{Fig3}
\end{figure}

In order to analyze the unfolding pathway of the peptide we proceeded in the following way.
Starting from the matrix built from all transitions among the clusters we performed a Perron cluster cluster analysis (PCCA$^+$)\cite{MSMbook} to coarse grain the MSM. 
We find that six eigenvalues of the transition matrix are larger than the inverse lagtime of 1 ns and therefore use 7 metastable states in the description of the thermal unfolding.
From the estimated mean first passage times for transitions among the metastable states the probability for various unfolding pathways is obtained.
The resulting thermal unfolding pathway is shown in Fig.\ref{Fig3} where the structures of the peptide for the various states are shown as superpositions along with the PMF.

It is clear that the thermal unfolding pathway dominantly proceeds via the opening of the N-terminus (positive values of IC$_2$) and only afterwards the complete unfolding takes place in a 'zipper-like' manner.
This pathway is comparable to the one observed earlier in water instead of methanol and under the influence of small constant external forces\cite{Knoch:2017}.
Some misfolded structures are additionally presented in Fig.\ref{Fig3} showing that the unfolding of this small peptide exhibits some degree of heterogeneity and cannot be characterized as a simple cooperative unfolding process.

\begin{figure}[h!]
\vspace{-0.25cm}
\centering
\includegraphics[width=9.0cm]{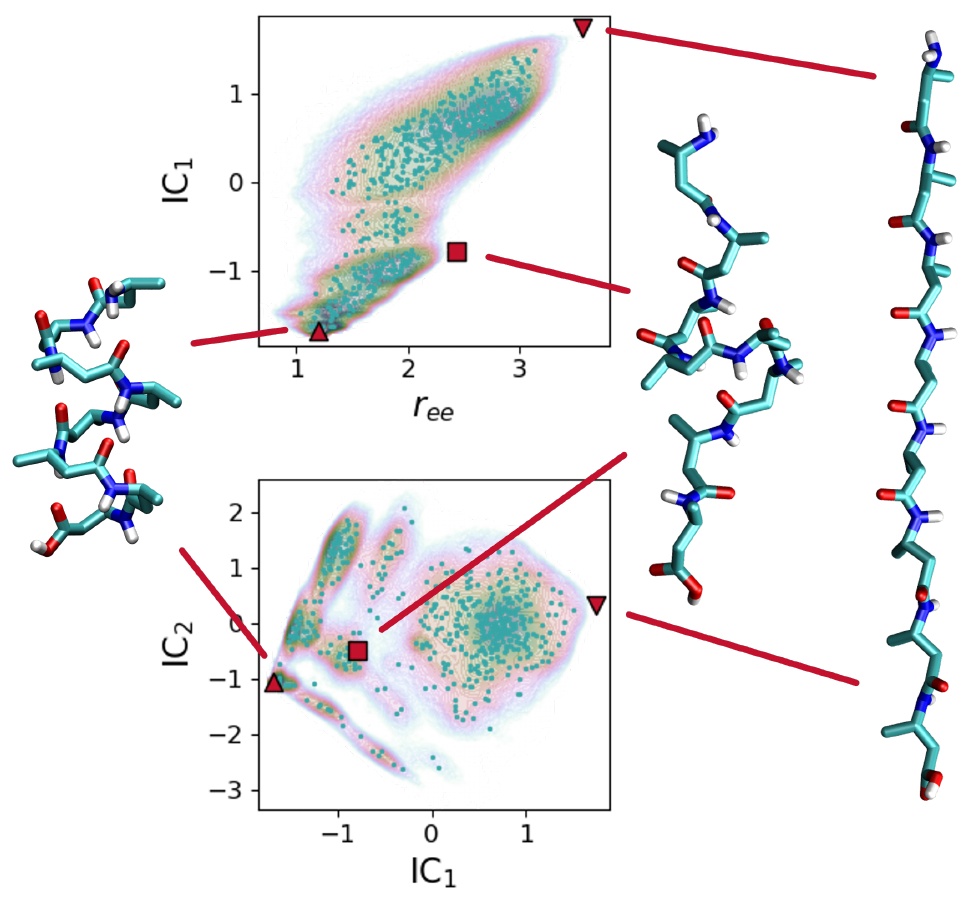}
\vspace{-0.5cm}
\caption{Potential of mean force obtained from an Umbrella sampling calculation at $T=240$ K using a force constant of 0.1 N/m.
The structures shown are those also identified in the FPMD simulations, cf. Fig.\ref{Fig1}.
}
\label{Fig4}
\end{figure}

For the reconstruction of a force ramp protocol we performed an Umbrella Sampling (US)\cite{Kastner:2011} calculation at $T=240$ K.
The resulting PMF is shown in Fig.\ref{Fig4}.
A comparison to the PMF obtained from a MD simulation at $T=298$ K in Fig.\ref{Fig2} shows some similarities but also significant differences.
The mimimum located near the intermediate structure ($r_{ee}\sim2.4$ nm, IC$_1$$\sim-1$) is more pronounced in the US calculation than in the MD simulation and can be viewed as responsible for the stability of this structure in the FPMD simulations.
\subsubsection*{Markov state model for force ramp simulations}
\begin{figure}[h!]
\vspace{-0.25cm}
\centering
\includegraphics[width=13.0cm]{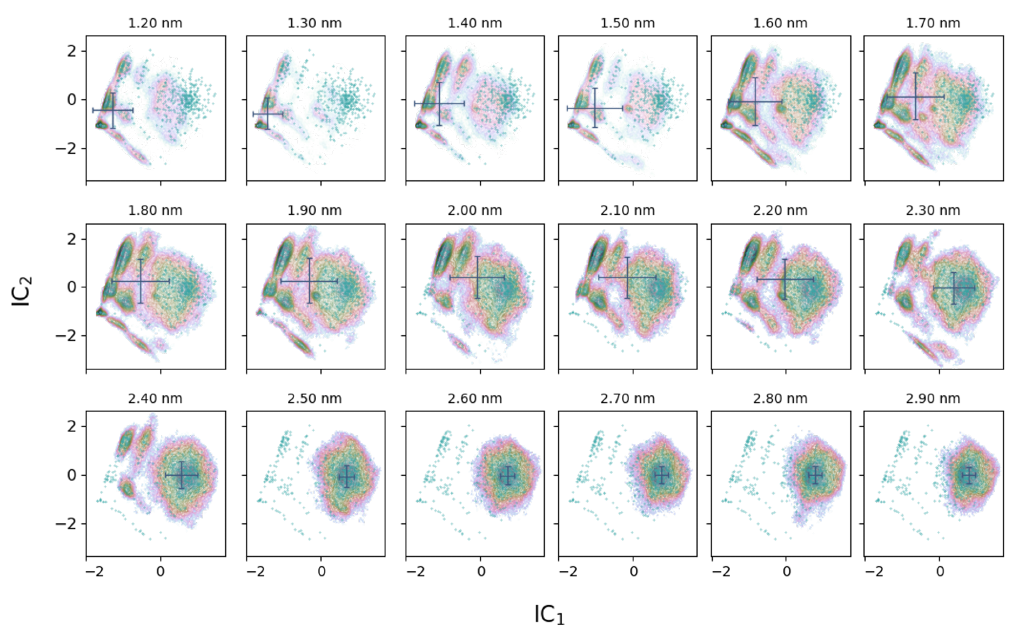}
\vspace{-0.25cm}
\caption{Potential of mean force obtained from Umbrella sampling calculations for different values of $r_s^{(m)}$ as indicated at 
$T=240$ K.
The light blue circles indicate the cluster centers.
The crosses indicate the center and the standard deviations of the probability distributions $P$(IC$_1$,IC$_2$).
}
\label{Fig5}
\end{figure}

As described in detail in ref.\cite{G89}, the dynamic coarse graining procedure is set up in the following way.
The sampling coordinate $r_s$ is discretized into 21 equally spaced values $r_s^{(m)}$, $m\!=\!1, 2, \cdots\!, M\!=\!21$, in the interval from 1.2 nm to 3.2 nm, defining $M$ ensembles.
In order to compute the PMF for each ensemble '$m$' and to sample transitions between neighboring ensembles, i.e. between '$(m-1)$', '$m$', and '$(m+1)$', we proceeded in the following way.
We performed simulations with a constant pulling velocity of 0.1 m/s until $r_s$ reached values $r_s^{(m-1)}$, $r_s^{(m)}$, and $r_s^{(m+1)}$. 
Afterwards, we applied a biasing potential given by $U^{(m)}=(K/2)(r_s^{(m)}-r_{ee})^2$, $K=0.1$ N/m, to each of these ensembles and carried out 10 simulations of 500 ns duration each.
This allows an effective sampling of the transitions within one ensemble and between different ensembles. 
Examples of the resulting PMFs are presented in Fig.\ref{Fig5}.
As expected, from these PMFs the unfolding of the peptide can be observed from the increasing value for IC$_1$ with increasing $r_s$.
Furthermore, it is apparent that particularly for intermediate values of $r_s$ more than two minima exist indicating the force dependent complexity of the energy landscape that is hard to be captured using a single order parameter.

For each of the 'm'-ensembles, a MSM in four dimensions is set up using IC$_1$, IC$_2$, IC$_3$, $r_{ee}$ as CVs, 
CV$_1$,$\cdots$,CV$_4$.
The states of the MSM are defined via the cluster centers  $\a=1,\cdots,N_{Cl}=500$ and the transition rate matrix ${\bf W}^{(m)}$ of all transitions among the CVs
CV$_{k}^{\b}\to$CV$_{l}^{\a}$ is determined from corresponding transition probabilities using the TRAM estimator with a lag time of 
1.25 ns\cite{G89,Wu:2016}.
The populations of the states are then given by ${\bf P}^{(m)}(t)=\exp{({\bf W}^{(m)}t)}{\bf P}^{(m)}(0)$
with the column vector ${\bf P}^{(m)}(t)=(P^{(m)}_1(t),\cdots,P^{(m)}_\a(t),\cdots, P^{(m)}_{N_{Cl}}(t))^T$, $m=1,\cdots,21$.

The force ramp mode implies a particular time dependence of the rate matrix ${\bf W}(t)$ that is represented by the sequence of 
${\bf W}^{(m)}$ with transitions between the ensembles in our approach.
This means that the system resides in the ensemble $m$ for a time interval $t\in(t_{m-1},t_m)$ and afterwards changes from $m$ to 
$(m+1)$ and the populations are given by
\[
{\bf P}^{(m)}(t)=e^{{\bf W}^{(m)}t}\!\cdot\!{\bf P}^{(m-1)}(t_{m-1}) \quad\mbox{with}\quad t\in(t_{m-1},t_m)
\]
because the system occupied the ensemble $(m-1)$ for times before $t_{m-1}$, i.e.
${\bf P}^{(m)}(0)={\bf P}^{(m-1)}(t_{m-1})$.
From the populations all observables of interest can be computed and in particular one has for the force:
\be\label{F.mittel}
\lg F(t)\rg = K\left[v\cdot t- (\lg r_{ee}(t)\rg - \lg r_{ee}(0)\rg)\right]\quad\mbox{with}\quad
\lg r_{ee}(t)\rg = \sum_{\a=1}^N r_{ee}^{(\a)}P_\a(t)
\ee
where $P_\a(t)$ is given by $P^{(m)}_\a(t)$ for $t\in(t_{m-1},t_m)$ and $\lg r_{ee}(0)\rg=r_s^{(0)}$ holds. 

\begin{figure}[h!]
\vspace{-0.25cm}
\centering
\includegraphics[width=14.0cm]{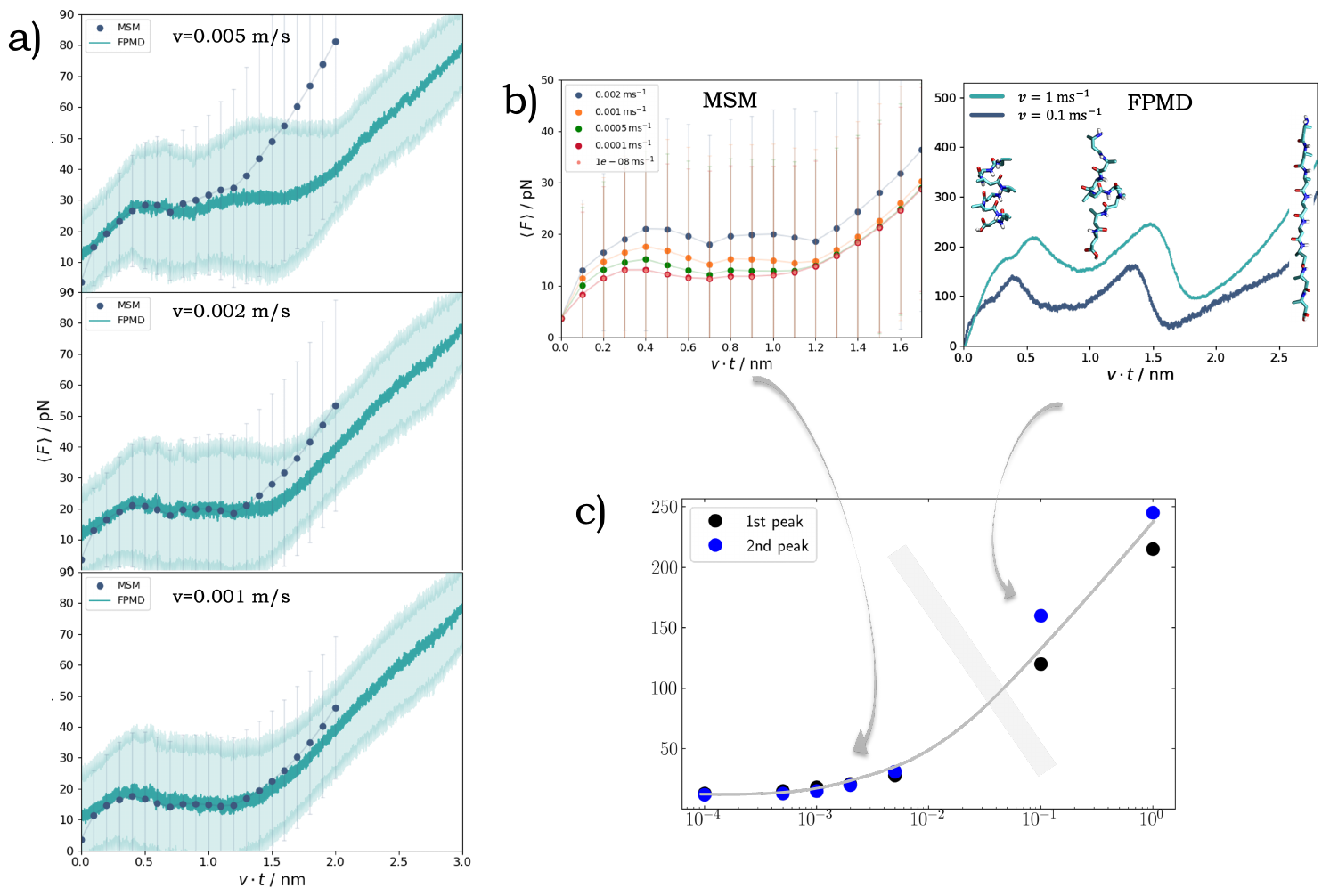}
\vspace{-0.5cm}
\caption{a): Force versus extension curves from Markov state modeling and from FPMD simulations at $T=240$ K. 
The shaded areas indicate the widths of the force distributions from 300 simulations. 
\newline
b): Comparison of mean forces obtained from Markov state modeling for small pulling velocities $v$ and those obtained from fast pulling FPMD simulations.
(1000 simulations for $v=1$ m/s and 300 simulations for $v=0.1$ m/s.)
\newline
c): Positions of the force peaks versus pulling velocity. The lines are guides to the eye.
}
\label{Fig6}
\end{figure}

Some results are presented in Fig.\ref{Fig6}. 
In Fig.\ref {Fig6}a) the results for the FECs of both methods, FPMD simulations and the MSM approach, are presented for the same set of parameters.
It is evident that the MSM calculations fail to reproduce the forces for the larger pulling velocities for extensions exceeding about 1 nm.
In this situation the MSM is likely to overestimate the mean force.
This is because for larger values of $r_s^{(m)}$, i.e. larger $v\cdot t$, the switch from the $m$-ensemble to the $(m+1)$-ensemble takes place faster than the transition rates from states $\a$ with small $r_{ee}^{(\a)}$ to states $\b$ with 
$r_{ee}^{(\b)}>r_{ee}^{(\a)}$.
Consequently, $\lg r_{ee}(t)\rg$ that is subtracted from the linear increase $(v\cdot t+\lg r_{ee}(0)\rg)$ in eq.(\ref{F.mittel}) is too small to give the correct value of the force.
In principle, the onset of this effect could be shifted to larger $v$ by using more ensembles in the construction of the TRAM.
However, there is another aspect that makes the MSM approach questionable at very high pulling speeds.
If the transitions among the ensembles becomes faster than the lag time used to construct the MSMs the Markovian character of the kinetics is no longer guaranteed.
For moderate and small pulling velocities the MSM approach can be used without further modification and gives reliable results.
In Fig.\ref{Fig6}b), the FECs for slow pulling (MSM) and fast pulling (FPMD) are shown.
It is evident that the overall structure, namley the occurence of two peaks, is similar in both cases.
The main difference is that the characteristic forces for fast pulling are almost one order of magnitude larger than for slow pulling.
Furthermore, both force maxima are more pronounced in the first case.
The results for $v=10^{-4}$ m/s and $v=10^{-8}$ m/s (not shown in Fig.\ref{Fig6}c)) are not distinguishable, cf. Fig.\ref{Fig6}a).
This indicates that the system reaches a state of quasi-equlibrium for very small pulling velocities\cite{Seifert:2002,G67}.
Furthermore, the maximum that for fast pulling is associated with the opening of the intermediate structure becomes extremely flat for small $v$.
This is compatible with the finding that in equilibrium the peptide opens starting at the N-terminus without a significant occurence of intermediate structures.
The values of the force maxima are plotted as a function of the pulling velocity in Fig.\ref{Fig6}c). 
Starting from very small $v$, after being almost constant, the force increases approximately linearly compatible with the phenomenological Bell model\cite{Dudko:2008,Evans:2001} and for larger $v$ a strong increase of the slope is observed. 
This behavior of the mean force as a function of the pulling velocity is well known and might indicate a change in the mechanism of the unfolding process\cite{Rief:2002}.
\section*{III. Conclusions}
In the present work we have extended the dynamic coarse graining method for FPMD simulations that we developed earlier for a model system that is well described by two states.
In that case a multi-ensemble Markov state modeling approach using the end-to-end distance of the system as an order parameter could be used to mimick a force-ramp protocol matching the pulling velocity scale of simulations and experimental force spectroscopy.
However, when the end-to-end distance is not a sufficient order parameter, the situation is more complex since the kinetics cannot be viewed as Markovian if a one-dimensional model is used.

In the present work, we considered a small $\b$-peptide and we used the H-bonds stabilizing the helical conformation as features for which we performed a dimension reduction using TICA.
If the three independent components with the largest variance are used in addition to the end-to-end distance, the Markovian approximation turned out to be valid.
In accord with earlier investigations we found that in equilibrium at $T=298$ K the peptide unfolds starting with the opening of the H-bonds located at the N-terminus followed by the consecutive opening of the H-bonds along the helix.
At lower temperatures ($T=240$ K) the helix is stable and the peptide can be unfolded by the application of an external force.
In earlier FPMD simulations it was found that the mechanical unfolding proceeds via a metastable intermediate which is characterized by open N- and C-termini and one helical loop in the center still intact.

We applied a Markov state modeling approach using the three dominant independent components and the end-to-end distance of the peptide as the relevant order parameters that defined the states of the system.
The pulling protocol was then mimicked by projecting the kinetics on the end-to-end distance and using this information to compute the measured forces.
For pulling velocities not too large the results match with those obtained with FPMD simulations using the same parameters. 
Our method allows to compute the observables of interest in the interesting regime of slow pulling where FPMD simulations are not feasible. 
In the chosen example, we can even reach the quasi-equilibrium behavior at extremely slow pulling speeds.
In summary, we have shown that the dynamic coarse graining technique developed earlier can be extended to systems exhibiting a complex unfolding pathway where a simple order parameter does not suffice to describe the kinetics.
\section*{Acknowledgement}
This project was funded by the Deutsche Forschungsgemeinschaft (DFG, German Research Foundation) in the framework to the collaborative research center "Multiscale Simulation Methods for Soft-Matter Systems" (TRR 146) under Project No. 233630050 (Project No. B3).
The authors  acknowledge the computing time granted on the supercomputer Mogon at Johannes Gutenberg University Mainz 
(hpc.uni-mainz.de).

\begin{thebibliography}{10}

\bibitem{Bieri:1999}
O.~Bieri and T.~Kiefhaber,
\newblock Biol. Chem. {\bf 380}, 923 (1999).

\bibitem{Woodside:2014}
M.~T. Woodside and S.~M. Block,
\newblock Annu. Rev. Biophys. {\bf 43}, 19 (2014).

\bibitem{Bustamante:2020}
C.~Bustamante, L.~Alexander, K.~Maciuba, and C.~M. Kaiser,
\newblock Ann. Rev. Biochem. {\bf 89}, 443 (2020).

\bibitem{Li:2022}
Q.~Li, D.~Apostolidou, and P.~E. Marszalek,
\newblock Methods {\bf 197}, 39 (2022).

\bibitem{Hummer:2003}
G.~Hummer and A.~Szabo,
\newblock Biophys. J. {\bf 85}, 5 (2003).

\bibitem{Dudko:2006}
O.~K. Dudko, G.~Hummer, and A.~Szabo,
\newblock Phys. Rev. Lett. {\bf 96}, 108101 (2006).

\bibitem{Dudko:2008}
O.~K. Dudko, G.~Hummer, and A.~Szabo,
\newblock Proc. Nat. Acad. Sci. USA {\bf 105}, 15755 (2008).

\bibitem{Lee:2009}
E.~H. Lee, J.~Hsin, M.~Sotomayor, G.~Comellas, and K.~Schulten,
\newblock Structure {\bf 17}, 1295 (2009).

\bibitem{Luitz:2015}
M.~Luitz, R.~Bomblies, K.~Ostermeir, and M.~Zacharias,
\newblock J. Phys.-Condens. Matt. {\bf 27}, 323101 (2015).

\bibitem{Perilla:2015}
J.~R. Perilla, B.~C. Goh, C.~K. Cassidy, B.~Liu, R.~C. Bernardi, T.~Rudack,
  H.~Yu, Z.~Wu, and K.~Schulten,
\newblock Curr. Opin. Struct. Biol. {\bf 31}, 64 (2015).

\bibitem{Isralewitz:2001}
B.~Isralewitz, M.~Gao, and K.~Schulten,
\newblock Curr. Opin. Struc. Biol. {\bf 11}, 224 (2001).

\bibitem{Park:2004}
S.~Park and K.~Schulten,
\newblock J. Chem. Phys. {\bf 120}, 5946 (2004).

\bibitem{Chen:2011}
L.~Y. Chen,
\newblock Phys. Chem. Chem. Phys. {\bf 13}, 6176 (2011).

\bibitem{Rico:2019}
F.~Rico, A.~Russek, L.~Gonzalez, H.~Grubm{\"u}ller, and S.~Scheuring,
\newblock Proc. Nat. Acad. Sci. {\bf 116}, 6594 (2019).

\bibitem{Valotteau:2019}
C.~Valotteau, F.~Sumbul, and F.~Rico,
\newblock Biophys. Rev. {\bf 11}, 689 (2019).

\bibitem{Franz:2020}
F.~Franz, C.~Daday, and F.~Gr\"ater,
\newblock Curr. Op. Struct. Biol. {\bf 61}, 132 (2020).

\bibitem{Best:2008}
R.~B. Best and G.~Hummer,
\newblock J. Am. Chem. Soc. {\bf 130}, 3706 (2008).

\bibitem{Habibi:2016}
M.~Habibi, J.~Rottler, and S.~S. Plotkin,
\newblock Plos Comput. Biol. {\bf 12}, e1005211 (2016).

\bibitem{Mucksch:2016}
C.~M\"ucksch and H.~M. Urbassek,
\newblock J. Chem. Theory Comput. {\bf 12}, 1380  (2016).

\bibitem{Ghosh:2017}
S.~Ghosh, A.~Chatterjee, and S.~Bhattacharya,
\newblock J. Chem. Theory Comput. {\bf 13}, 957 (2017).

\bibitem{G89}
F.~Knoch, K.~Sch{\"a}fer, G.~Diezemann, and T.~Speck,
\newblock J. Chem. Phys. {\bf 148}, 044109 (2018).

\bibitem{Wu:2016}
H.~Wu, F.~Paul, C.~Wehmeyer, and F.~No{\'e},
\newblock Proc. Nat. Acad. Sci USA {\bf 113}, E3221 (2016).

\bibitem{Knoch:2017}
F.~Knoch and T.~Speck,
\newblock Mol. Syst. Des. Eng. {\bf 3}, 204  (2017).

\bibitem{G84}
L.~Uribe, J.~Gauss, and G.~Diezemann,
\newblock J. Phys. Chem. B {\bf 119}, 8313 (2015).

\bibitem{Cheng:2001}
R.~P. Cheng, S.~H. Gellman, and W.~F. DeGrado,
\newblock Chem. Rev. {\bf 101}, 3219  (2001).

\bibitem{G85}
L.~Uribe, J.~Gauss, and G.~Diezemann,
\newblock J. Phys. Chem. B {\bf 120}, 10433 (2016).

\bibitem{Praprotnik:2006}
M.~Praprotnik, L.~Delle~Site, and K.~Kremer,
\newblock Phys. Rev. E {\bf 73}, 066701 (2006).

\bibitem{G104}
M.~Oestereich, J.~Gauss, and G.~Diezemann,
\newblock J. Chem. Phys. {\bf 161}, 154903 (2024).

\bibitem{Seebach:1997}
D.~Seebach and J.~L. Matthews,
\newblock Chem. Commun. , 2015  (1997).

\bibitem{Gademann:1999}
K.~Gademann, B.~Jaun, D.~Seebach, R.~Perozzo, L.~Scapozza, and G.~Folkers,
\newblock Helv. Chim. Acta {\bf 82}, 1  (1999).

\bibitem{Abraham:2015}
M.~J. Abraham, T.~Murtola, R.~Schulz, S.~Pall, J.~C. Smith, B.~Hess, and L.~E.,
SoftwareX {\bf 1 -2}, 19 (2015).

\bibitem{Goga:2012}
N.~Goga, A.~J. Rzepiela, A.~H. de~Vries, S.~J. Marrink, and H.~J.~C. Berendsen,
\newblock J. Chem. Theory Comput. {\bf 8}, 3637 (2012).

\bibitem{Oostenbrink:2004}
C.~Oostenbrink, A.~Villa, A.~Mark, and W.~van Gunsteren,
\newblock J. Comput. Chem. {\bf 25}, 1656 (2004).

\bibitem{Walser:2000}
R.~Walser, A.~E. Mark, W.~F.~v. Gunsteren, M.~Lauterbach, and G.~Wipff,
\newblock J. Chem. Phys. {\bf 112}, 10450 (2000).

\bibitem{ewaldsum}
T.~Darden, D.~York, and L.~Pedersen,
\newblock J. Chem. Phys. {\bf 98}, 10089 (1993).

\bibitem{ewaldsum2}
U.~Essmann, L.~Perera, and M.~Berkowitz,
\newblock J. Chem. Phys. {\bf 103}, 8577 (1995).

\bibitem{Hess:1997}
B.~Hess, H.~Bekker, H.~Berendsen, and J.~Fraajie,
\newblock J. Comput. Phys. {\bf 18}, 1463 (1997).

\bibitem{Allen:1987}
M.~Allen and D.~Tildesley,
\newblock {\em Computer Simulations of Liquids},
\newblock Oxford, Oxford Science Publications, 1987.

\bibitem{Abraham:2018}
M.~Abraham, D.~van~der Spoel, E.~Lindahl, and B.~Hess,
\newblock www.gromacs.org  (2018).

\bibitem{Berendsen:1984}
H.~J.~C. Berendsen, J.~P.~M. Postma, W.~F. van Gunsteren, A.~DiNola, and J.~R.
  Haak,
\newblock J. Chem. Phys. {\bf 81}, 3684 (1984).

\bibitem{Noe:2017}
F.~No{\'e} and C.~Clementi,
\newblock Curr. Opin. Struct. Biol {\bf 43}, 141  (2017).

\bibitem{Bhakat:2022}
S.~Bhakat,
\newblock RSC Advances {\bf 12}, 25010 (2022).

\bibitem{Molgedey:1994}
L.~Molgedey and H.~G. Schuster,
\newblock Phys. Rev. Lett. {\bf 72}, 3634 (1994).

\bibitem{PerezHernandez:2013}
G.~Perez-Hernandez, F.~Paul, T.~Giorgino, G.~D. Fabritiis, and F.~No{\'e},
\newblock J. Chem. Phys. {\bf 139}, 015102 (2013).

\bibitem{Noe:2015}
F.~No{\'e} and C.~Clementi,
\newblock J. Chem. Theory Comput. {\bf 11}, 5002  (2015).

\bibitem{MSMbook}
G.~R. Bowman, V.~S. Pande, and F.~No{\'e},
\newblock {\em An introduction to Markov state models and their application to
  long timescale molecular simulation}, volume 797,
\newblock Springer Science \& Business Media, 2013.

\bibitem{Scherer:2015}
M.~K. Scherer, B.~Trendelkamp-Schroer, F.~Paul, G.~Perez-Hernandez, M.~Hoffmann, N.~Plattner, C.~Wehmeyer, J.-H. Prinz, and F.~No{\'e},
\newblock J. Chem. Theory Comput. {\bf 11}, 5525  (2015).

\bibitem{Kastner:2011}
J.~K\"astner,
\newblock WIREs Comput. Mol. Sci. {\bf 1}, 932  (2011).

\bibitem{Seifert:2002}
U.~Seifert,
\newblock Europhys. Lett. {\bf 58}, 792 (2002).

\bibitem{G67}
G.~Diezemann and A.~Janshoff,
\newblock J. Chem. Phys. {\bf 129}, 084904 (2008).

\bibitem{Evans:2001}
E.~Evans,
\newblock Annu. Rev. Biophys. Biomol. Struct. {\bf 30}, 105 (2001).

\bibitem{Rief:2002}
M.~Rief and H.~Grubm\"uller,
\newblock ChemPhysChem. {\bf 3}, 255 (2002).

\end{thebibliography}
\end{document}